\documentclass[amsmath,amssymb,twocolumn,nofootinbib,floatfix]{revtex4-2}
\usepackage{pdfpages}
\makeatletter
\AtBeginDocument{\let\LS@rot\@undefined}
\makeatother

\usepackage{color}
\usepackage[colorlinks=true,linkcolor=blue,urlcolor=blue,citecolor=blue]{hyperref}
\usepackage{graphicx}\usepackage{bm,color}

\usepackage{float}
\usepackage[caption = false]{subfig}

\definecolor{labelkey}{cmyk}{.4,.2,0,0}
\usepackage{times}

\newcommand{\highlight}[1]{\colorbox{yellow!30}{\rule[-1.5ex]{0mm}{4.5ex}{\mbox{~$\displaystyle #1$~}}}}
\newlength{\highlightlength}
\newlength{\highlightlengthbis}

\newcommand{\hl}[1]{{\highlight{#1}}}

\usepackage{tikz}
\usetikzlibrary{arrows,shapes,positioning}
\usetikzlibrary{decorations.markings}
\tikzstyle arrowstyle=[scale=1]
\tikzstyle directed=[postaction={decorate,decoration={markings,
    mark=at position .65 with {\arrow[arrowstyle]{stealth}}}}]
\tikzstyle endreversedirected=[postaction={decorate,decoration={markings,
    mark=at position 1.0 with {\arrow[arrowstyle]{stealth}}}}]
\tikzstyle enddirected=[postaction={decorate,decoration={markings,
    mark=at position 1.0 with {\arrow[arrowstyle]{stealth}}}}]

\tikzstyle reverse directed=[postaction={decorate,decoration={markings,
    mark=at position .65 with {\arrowreversed[arrowstyle]{stealth};}}}]

\newcommand{\Eq}[1]{Eq.~(\ref{#1})}

\newcommand{\eq}[1]{(\ref{#1})}
\newcommand{\ds}[1]{\displaystyle }

\newcommand{\bea}{\begin{eqnarray}}
\newcommand{\eea}{\end{eqnarray}}
\newcommand{\beq}{\begin{equation}}
\newcommand{\eeq}{\end{equation}}

\definecolor{mygrey}{gray}{0.5}

\definecolor{mygray}{gray}{0.4}

\newcommand{\rme}{\mathrm{e}}
\newcommand{\rmd}{\mathrm{d}}
\newcommand{\nn}{\nonumber}
\renewcommand{\epsilon}{\varepsilon}
\newcommand{\nott}[1]{}
\newcommand{\ca}[1]{{\cal #1}}
\newcommand{\be}{\begin{equation}}
\newcommand{\ee}{\end{equation}}
\newcommand{\Fig}[1]{\includegraphics[width=\columnwidth]{./#1}} 
\newcommand{\fig}[2]{\includegraphics[width=#1\columnwidth]{./#2}}

\newlength{\bilderlength}

\graphicspath{{figures/},{}}

\begin{document}

\title{Locating the Ising CFT via the  ground-state energy on the fuzzy sphere}
\author{Kay J\"org Wiese}
  \affiliation{CNRS-Laboratoire de Physique de l'Ecole Normale Sup\'erieure, PSL Research University, Sorbonne Universit\'e, Universit\'e Paris Cit\'e, 24 rue Lhomond, 75005 Paris, France.}

\begin{abstract}

We  locate  the  phase-transition line for the   Ising model on the fuzzy sphere from a finite-size scaling analysis of its ground-state energy.  Our strategy   is to write 
the latter as $E_{\rm GS}(N_m)/N_m = E_{0} + E_1 /N_m  + E_{3/2}/N_m^{3/2}+  ...$, and to search for a minimum of $ \chi:=E_{3/2}/E_0$ as a function of the couplings.
Conformal perturbation theory predicts that around a CFT, 
$\chi= \chi_{\rm min} + \sum_i \lambda_i^2 N_m^{- \omega_i} + \ca O(\lambda^3)$, where  $\lambda_i$ are the couplings associated to perturbations of operators with dimension $\Delta_i$, and $\omega_i = d-\Delta_i$.
This procedure finds the critical curve of [\href{https://dx.doi.org/10.1103/PhysRevX.13.021009}{\rm PRX {\bf 13} (2023)
  021009}]     and their sweet spot with good precision.
Varying two coupling constants allows us to extract the correction-to-scaling exponent  $\omega$ associated to the two leading scalars $\epsilon$, and 
$\epsilon'$.   
We  find similar results when normalizing by the gap to the stress tensor $T$ or first parity-odd operator $\sigma$ instead of $E_0$.

\end{abstract} 

\maketitle

\section{Introduction}
The fuzzy-sphere regularization is an exciting new tool \cite{ZhuHanHuffmanHofmannHe2023} to access conformal field theories (CFTs) in 
dimension $d=2+1$. The technique uses a 2-sphere inside which are placed $s$ magnetic monopoles. 
The lowest Landau level on this sphere is   
$N_m=2s+1$ times degenerate, each with  $N_{\rm f}$ flavors ($N_{\rm f}=2$ for the Ising model).  
Filling $N_m$   of the $N_m\times N_{\rm f}$ states, and adding 
interactions between the electrons, allows one to engineer CFTs with a given symmetry. 
This approach  was introduced in \cite{ZhuHanHuffmanHofmannHe2023} to study the quantum 2d Ising model, equivalent to the classical 3d   model.  
It was since   applied to other systems, among which are theories with global
  $\mathrm{Sp}(n)$    \cite{ZhouHe2024} or $O(3)$ \cite{DeyHerviouMudryLauchli2025} symmetry, Majorana fermions \cite{ZhouGaiottoHe2025},
a free scalar \cite{He2025},  Chern-Simons matter  \cite{ZhouWangHe2025},   the Potts model \cite{YangYueTangHanZhuChen2025}, and Yang-Lee  \cite{FanDongVishwanath2025,CruzKlebanovTarnopolskyXin2025,MiroDelouche2025}. The latter example is interesting as   the  theory is non-unitary, even though its spectrum is real; thus it is not accessible via the numerical conformal bootstrap
\cite{ReehorstRychkovSimmons-DuffinSiroisSuRees2021,PolandRychkovVichi2019,El-ShowkPaulosPolandRychkovSimmons-DuffinVichi2014,El-ShowkPaulosPolandRychkovSimmons-DuffinVichi2012,ChesterLandryLiuPolandSimmons-DuffinSuVichi2019} which is the current gold standard for 3d CFTs. 

While the method allows one to study phases as well as critical points, the latter are particularly interesting, especially when they correspond to a CFT. 
A crucial step is to identify the location of the critical point. The standard approach consists in choosing 
couplings which optimize properties of the intended CFT, as the integer spacing of descendents \cite{ZhuHanHuffmanHofmannHe2023}. Doing so  without assuming what one wants to show is 
critical. There are interesting  recent proposals \cite{FardelliFitzpatrickKatz2024,Fan2024} to construct the conformal generators on the fuzzy sphere, allowing one to check for the overlap 
of states expected to be descendants, with the constructed descendants of expected primaries.

Here we  locate the  phase-transition line and sweet spot for the   Ising model on the fuzzy sphere from a finite-size scaling analysis of the ground-state (GS) energy, without using any information on higher excited states. 
Our procedure finds the critical curve of  \cite{ZhuHanHuffmanHofmannHe2023} with good precision, and their sweet spot  (point of optimal conformality) as well. 

\enlargethispage{1cm}

\section{Size-dependence of the ground state energy}
\label{Size-dependence of the ground state energy}
In dimension    $d=2$,
the free energy per site of a classical CFT on a torus of circumference $L$ is given by  \cite{BloteCardyNightingale1986,CardyLH1988}, 
\be\label{ceff}
f(L) = f_\infty -  \frac {\pi c} {6  L^2} + \ca O(L^{-3}).
\ee
Here $f_\infty(L)$ is the free energy per site in the limit of $L\to \infty$, and $c$ its central charge. 
This formula is   valid   at the critical point, and can be used to locate it: first one measures an effective central charge $c_{\rm eff}(g|L)$, where $g$ is a microscopic coupling, via a fit to  \Eq{ceff}; then one    demands that the theory is at a marginal point  for $c_{\rm eff}(g|L)$\cite{JacobsenWiese2024}, 
\be\label{der-ceff}
\partial _g c_{\rm eff}(g|L) \big|_{g=g^*} = 0\quad \Longrightarrow \quad\lim_{L\to \infty}   c_{\rm eff}(g^*|L) = c. 
\ee
For a standard (real) CFT, a local maximum for $c_{\rm eff}(g|L)$ indicates a critical point, while a minimum indicates a critical phase. 
If the  model at $g^*$ is a CFT, then  $c_{\rm eff}(g^*|L)$ converges for $L\to \infty$ to the central charge of the corresponding CFT. 

The central charge has three properties:
{({\em i})} it appears in the Virasoro algebra of conformal generators, {({\em ii})} it governs   finite-size corrections as in \Eq{ceff}, and it decreases along the RG flow (Zamolodchikov's $c$-theorem \cite{Zamolodchikov1986}). This motivates the condition \eq{der-ceff} for  fixed points,  and explains why a critical point is a local maximum. 
 The marginality condition \eq{ceff} can also be used for complex CFTs, and allowed the authors of  \cite{JacobsenWiese2024} to locate the critical point of the 5-state Potts model in $d=2$, which has a complex second derivative, thus is neither maximum nor minimum.

There is a generalization of Zamolodchikov's $c$-theorem to $d=4$,   related to anomalies under 
Weyl rescaling 
\cite{DeserSchwimmer1993,Duff1994}, which gives  the stress-energy tensor in the form (see Eq.~(1.2) of \cite{KomargodskiSchwimmer2011})
\be
T_\mu^{\mu} = a E_4 - c W_{\mu \nu \rho \sigma}^2 .
\ee
Here $E_4$ is the Euler density (which integrates to the Euler characteristic), and $W$ the Weyl tensor. 
It was   first conjectured and checked to 1-loop order by Cardy \cite{Cardy1988}, and later    proven  in \cite{KomargodskiSchwimmer2011} that $a$ decays along RG trajectories ($a$-theorem).

We now consider  dimension $d=3$. Ref.~\cite{JafferisKlebanovPufuSafdi2011} (section 1) considers finite-size corrections for the free energy of an Euclidean field theory on a  3-sphere of radius $R$ imbedded into $\mathbb R^4$, 
\be\label{4}
 F = - \ln |Z_{S^3}|  = \alpha_0 R^3 + \alpha_1 R + \ca F . 
\ee
While  $\alpha_0$ and $\alpha_1$ are non-universal, the last term $\ca F$ is a number, and may contain universal information. 
$\ca F$ was first obtained  via holography \cite{MyersSinha2011} and in super-symmetric models \cite{Jafferis2012},  and only later for 
more general theories \cite{KlebanovPufuSafdi2011}. 
It took some time to find a proper definition   which is universal, and independent of the microscopic degrees of freedom 
\cite{HertzbergWilczek2011,LiuMezei2013}.
Currently, the best  approach to extract  $\ca F$ is to study the entanglement spectrum  \cite{CasiniHuerta2012}. 
Refs.~\cite{CasiniHuerta2012,LiuMezei2013}
proved that   $\ca F$ so defined descends along the RG flow. 
 Ref.~\cite{HuZhuHe2024} succeeded to extract $\ca F$ from  the entanglement entropy on the fuzzy-sphere realization of the Ising model.

Now consider the free energy on $\mathbb R^3 \simeq \ca S_2 \times \mathbb R$, where the r.h.s.\ is the geometry used in the fuzzy-sphere 
approach. To pass to the left, Euclidean time $\tau$ is mapped as  $\tau\to \rme^\tau$, with $\tau \in [-\beta/2,\beta/2]$. For   $\beta\gg R$, we expect the free energy  to be 
\be
F = - \ln Z = \beta \left[ \alpha_0 R^2 + \alpha_1 + \frac {\alpha_{3/2}}{R} + ...\right].
\ee
The coefficient which may contain universal information is $\alpha_{3/2}$, as it is the only one without a scale. 
We can pass to the quantum problem by taking $\beta\to \infty$,  
\be\label{EGS}
E_{\rm GS}(R) = v_{\rm F}\left[ \alpha_0 R^2 + \alpha_1 + \frac {\alpha_{3/2}}{R} + ...\right]. 
\ee
Here $v_{\rm F}$ is un unknown Fermi velocity, which in general depends on the microscopic couplings. 

Let us explain why we did not include a term of order $R$ into $E_{\rm GS}(R)$:  the two leading terms in the vacuum energy allowed by geometry and necessitating counter-terms (CT) are
\be\label{7}
E_{\rm GS}^{\text{CT}}(R) = \int \rmd^2x\, \sqrt{g} \Big(  \mu  + \kappa  \ca R \Big),
\ee
where $R$ is the radius of the sphere, $\rmd^2x \sqrt{g}$  the invariant measure,  $ \mu$    the coefficient of the  vacuum-energy density (``cosmological constant''), and $ \kappa$ the coefficient multiplying  the  curvature  $\ca R$.  According to the Gau\ss-Bonnet theorem, the latter integrates to $4 \pi \chi$, where $\chi $ is the Euler characteristic; $\chi=2$ for a sphere.  As a result $E_{\rm GS}^{\text{CT}}(R) =  4 \pi R^2 \mu + 8 \pi \kappa $.

Using the identification that the area of the sphere grows as the number of electrons, or more precisely \cite{Zhou2025}
\be\label{Nm-R}
N_m^2 = 1 + 4 R^4, 
\ee
we get for 
the GS energy per electron\footnote{\label{FuzzifiED-footnote}The GS energy given by FuzzifiED is indeed   $\sim N_m$.}    
\be\label{ansatz}
\frac{E_{\rm GS}(N_m)}{N_m} = E_{0} + \frac{E_1}{N_m}  + \frac{E_{3/2}}{N_m^{3/2}}+ ... 
\ee  
We recall that   physical information  is expected in $E_{3/2}$. 
This is in line with the observation that the gap of any excited state  has the same scaling in $N_m$ as $E_{3/2}$ 
(for the stress-energy  gap $E_T-E_{\rm GS}$ see discussion below and Fig.~\ref{f:ETminusEGSoverEGS}); thus if one looks for a   coefficient of the GS energy which is universal, one again concludes that  the only viable  candidate is $E_{3/2}$.

At this point it is interesting to recall a standard tool in   data analysis, the Binder cumulant a.k.a.\ kurtosis (ratio of fourth moment divided by the square of the second moment). 
In the vicinity of a phase transition, it takes the form
\be\label{Binder}
\ca B(\{\lambda_i\}|R) = \ca B_0 +p ( \{\lambda_i R^{-\omega_i}\}),\qquad \omega_i = \Delta_i-d,
\ee
where $\ca B_0$ is a number, and $p$ a function\footnote{When a perturbative expansion in $\lambda_i$ exists, $p$ is a polynomial.} in the couplings $\lambda_i$ associated with perturbations by operators of dimension $\Delta_i$. 
We postulate that $E_{3/2}$ can play a similar role to the Binder cumulant  in locating  the point where finite-size corrections are minimized. 

Now consider the behavior close to a CFT. There  the action takes the form
\be
\ca S = \ca S^{\rm CFT}+ \sum _i \lambda_i \int_x \ca X_i(x),
\ee
where $\ca X_i(x)$ are primaries. 
Schematically, the free energy in   statistical field theory is \cite{LudwigCardy1987}
\bea
F &=& - \ln Z = F_0^{\rm CFT} + \sum_i \lambda_i \int_x \left< \ca X_i(x)\right>_{\rm CFT}\nn\\
&& - \frac12 \sum_{i,j} \int_{x_1}\int_{x2}\left< \ca X_i(x_1)\ca X_j(x_2)\right>^{\rm c}_{\rm CFT} + ...
\eea
As the vacuum expectation   of   primaries vanishes, the linear terms are absent. 
At second order, only the term with $i=j$ survives. The result is 
\be\label{13}
 F = F_0^{\rm CFT} -   \sum_i  \left( {\lambda_i}  R^{d-\Delta_i}\right)^2 . 
\ee
We have not specified our domain of integration, and   absorbed possible  geometrical factors into $\lambda_i$. 
\Eq{13} has a  factor of $R^{2d}$   from the integration measure, and a   factor of $R^{-2\Delta_i}$ from the  2-point function 
$\left< \ca X_i(0)\ca X_i(R)\right>= R^{-2\Delta_i}$. 
An equivalent procedure can   be performed in the quantum-formulation \cite{HogervorstRychkovRees2015,LaeuchliHerviouWilhelmRychkov2025} which uses the sphere as geometry\footnote{In this formulation, the integration measure gives $R^{2d-2}$. The seemingly missing factor of $R^2$ is provided by the denominator $E_i-E_{\rm GS}\sim R$, correcting an energy $\sim R$. The quantum version with its well-defined geometry is  appropriate  for evaluating the geometrical prefactors.}. 
We conclude that 
\be\label{14}
E_{ 3/2} = E_{ 3/2}^{\rm CFT} - \sum_i  \left( {\lambda_i}  R^{d-\Delta_i}\right)^2 + \ca O(\lambda_i^3). 
\ee
A cubic term is expected   as there are non-vanishing 3-point functions. 
\Eq{14} is of the  form postulated in \Eq{Binder}. 
In order to eliminate the Fermi-velocity $v_{\rm F}$ in \Eq{EGS}, below we shall analyze\footnote{If there are perturbative corrections to $E_0$, we expect them to take the same form as in \Eq{14}, so that the  argument remains valid for the ratio.}
\be\label{K(h,g)}
\hl{\chi   :=\frac{E_{3/2}}{E_0}.} 
\ee
(Alternative normalizations are discussed in section \ref{Alternative normalizations}.)
Since in our problem  $E_0$ is negative, a CFT implies  a minimum of $\chi$. Around this minimum, we expect
\be\label{chi-expansion}
\hl{\chi   = \chi_{\rm min} + \sum_i \frac{\lambda_i^2}{N_m^{\omega_i}}, \quad \omega_i = \Delta_i-d.}
\ee
As we   show below,  this procedure finds the critical curve and the sweet spot of  \cite{ZhuHanHuffmanHofmannHe2023} with good precision. 
 The standard algorithm to locate a CFT is to search for a spectrum with integer-valued spacing for   descendants  
of primary operators, as dictated by  CFT  \cite{ZhuHanHuffmanHofmannHe2023}, or explicitly check for conformal symmetry \cite{FardelliFitzpatrickKatz2024,Fan2024}.
Our procedure is an alternative. Its  sole ingredient is the GS energy. 
Its advantages are that it is simple to implement, and computationally  fast.

\section{Finite-size scaling analysis of the ground-state energy}
\subsection{Model}
We use the Ising model defined in the seminal work \cite{ZhuHanHuffmanHofmannHe2023}. It consists of a 2-sphere   with $s$ magnetic monopoles at its center,   onto which are placed $N_m=2s+1$   fermions, each with two internal degrees of freedom (up and down spin). Without interactions, the   spectrum is flat. One then introduces a repulsive 
interaction upon contact between electrons pointing in the $z$-direction of strength $1$, and a ``kinetic'' term (coupling to the gradient of the density interactions) of strength $g\equiv V_1$ in the same direction. Finally, a transverse magnetic field of strength $h$   drives the phase transition, similar to what happens in the 1-dimensional spin chain.

\subsection{FuzzifiED}

We use the package \href{https://docs.fuzzified.world}{FuzzifiED}. It provides a compact and   efficient implementation to obtain the spectrum of a user-defined 
model on the fuzzy sphere, both using exact diagonalization and DMRG. 
 The package is described in detail in \cite{Zhou2025}.
The program we used is an adaptation of  ``ising\underline{~}spectrum.jl'', wrapped inside a Mathematica loop. To avoid numerical errors, 
we  eliminated the rounding, i.e.\ the instruction   ``round.([enrg[i], l2\underline{~}val[i], P, Z], digits = 6)''. 
For system sizes $N_m \ge 16$ we restrict the evaluation to the GS energy, leading to a considerable speedup. 
(We can   do $N_m=17$ in about 10 minute on 8 cores, using  a machine from 2013.)

\begin{figure}[t]
\Fig{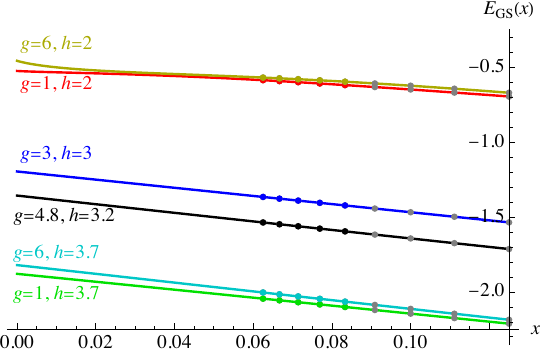}
\caption{$E_{\rm GS}(x)$ for various values of $h$ and $g$, for $o=4$, using $x^{i/2}$ as a basis, $\{ 1, x, x^{3/2},x^2,x^{5/2}\}$, $N_m^{\rm max}=16$. Gray dots are  not used for the fit, but in agreement with it.}
\label{f:F1b}
\end{figure}

\begin{figure}[t]
\Fig{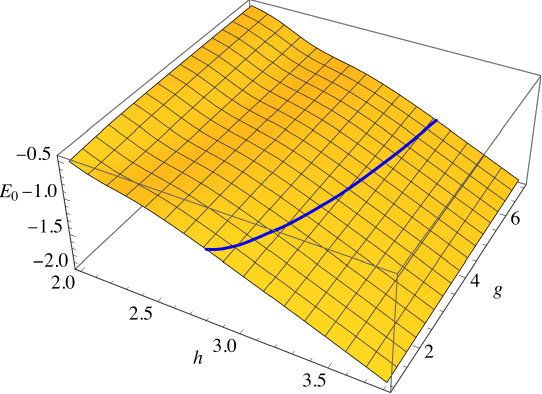}
\caption{$E_0= E_{\rm GS}(0|h,g)$ for $o=4$, $N_m^{\rm max}=16$; this plot changes little between  $N_m^{\rm max}=13$ and $N_m^{\rm max}=16$. In blue the critical line.}
\label{E0ofhg}
\end{figure}
\subsection{Dependence of the GS energy on system size}

We had motivated the ansatz \eq{ansatz} for the GS energy. In the rest of this work, we     use 
this ansatz  with two additional subleading terms, 
\bea \label{half-integer-fit}
\frac{E_{\rm GS}(x|h,g)}{N_m} &=& E_0(h,g) + E_1(h,g) x + E_{3/2}(h,g)x^{\frac 32} \nn\\
&& +   E_{2}(h,g)x^{2} +  E_{5/2}(h,g)x^{\frac 52} +... \\
x&:=&\frac 1{N_m} . 
\eea
Let us check this against exact diagonalization. 
Fig.~\ref{f:F1b} shows that  the GS${}^{\ref{FuzzifiED-footnote}}$ has both a constant $E_0$ and  a strong linear component $E_1$.  
We   tried adding a term of order $\sqrt x$ and found that its coefficient is either absent or very small. 
Out of  curiosity, we also tried an ansatz with   integer coefficients. We found that 
information about the   critical line can be extracted, but seems to be sitting in higher derivatives, which we do not find convincing, see appendix \ref{s:Alternative fitting schemes}.  
We also  played with different orders of the truncation. 
We   trust our analysis when higher-order coefficients are small. 
With its almost straight behavior, 
Fig.~\ref{f:F1b} (which uses   \Eq{half-integer-fit}) suggests that the leading  term in the expansion is indeed of order $x$, and not $\sqrt x$, or 
 that at least the   coefficient $\sim \sqrt x$ is  very small, and we discard it. 
 Adding terms beyond $E_{3/2}$ seems  helpful (see appendix \ref{s:Alternative fitting schemes}).

We now look at the   GS energy extrapolated to $x=0$. As Fig.~\ref{E0ofhg} attests  this is rather featureless.
\begin{figure}[t]
\Fig{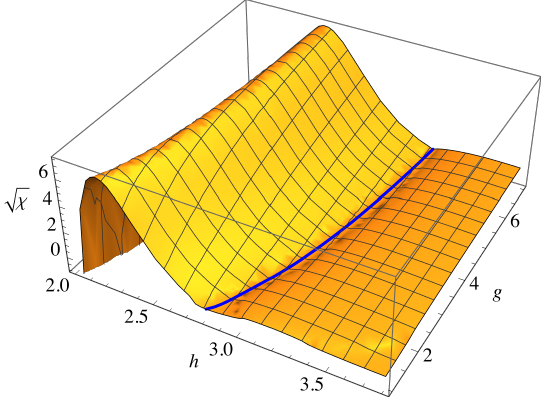}
\caption{$\chi $ for  $N_m^{\rm max}=16$; in blue the critical line. To make the minimum visible, we plot the signed   root ($\sqrt{\chi}:=\mbox{sign}(\chi) \sqrt{|\chi|}$). To enhance the resolution of the plot, the interpolation  outlined below in  \Eq{rho(h,g)} and Fig.~\ref{f:fit-weights} was used.}
\label{E3halfoverE-fitorder=4-nmMax=16SqrtFit}
\end{figure}
A crucial problem with   quantum-mechanical approaches  is that all energy levels are multiplied by an unknown Fermi velocity $v_F$, see \Eq{EGS}. In the fuzzy-sphere approach, this is usually  fixed by demanding that 
the stress-energy tensor have   dimension $d=3$. An alternative is to prescribe the energy of the first excited state $\sigma$   which suffers less from finite-size corrections close to the Ising CFT  \cite{LaeuchliHerviouWilhelmRychkov2025}. 
Our goal here is to extract the location of the CFT solely from the GS energy. To eliminate the unknown Fermi-velocity, we consider finite-size corrections   normalized by the extrapolated GS energy, i.e.
\be\label{K(h,g)-bis}
\chi  (h,g):=\frac{E_{3/2}(h,g)}{E_0(h,g)}, 
\ee
where the coefficients are those of \Eq{half-integer-fit}.  Alternative normalizations are discussed   in section \ref{Alternative normalizations}. 

\begin{figure}[t]
\Fig{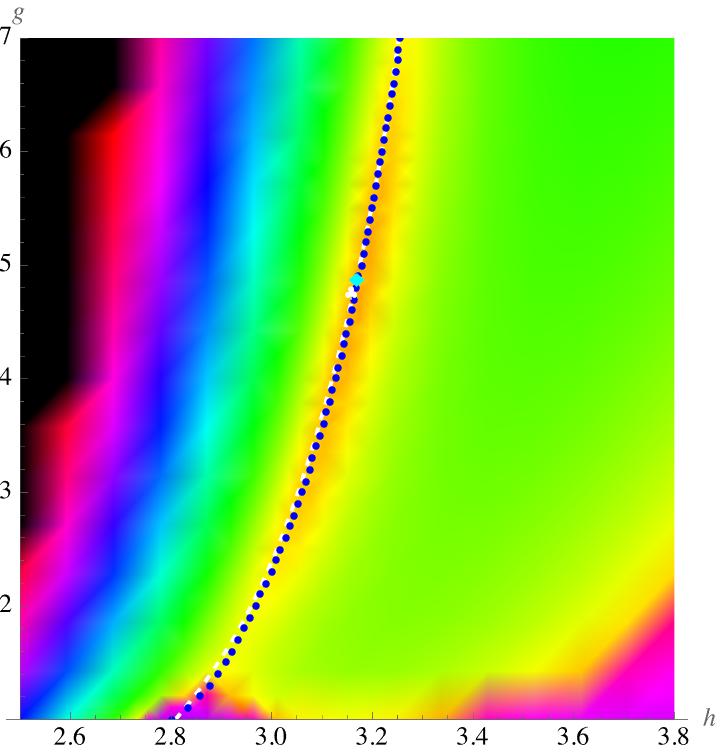}
\caption{Heat map of $\chi(h,g)$ given in \Eq{K(h,g)} for $o=4$, $N_m^{\rm max}=16$. The white dashed line is the critical line of \cite{ZhuHanHuffmanHofmannHe2023}, with a white shamrock marking their sweat spot (best agreement with a CFT). In dark blue dots the minimum of the valley of ${\chi (h,g)}$. The cyan diamond marks the global minimum of ${\chi (h,g)}$.}
\label{E3halfoverE-fitorder=4-nmMax=15SqrtFit-max-curvature-slope}
\end{figure}
\begin{figure}[t]
\Fig{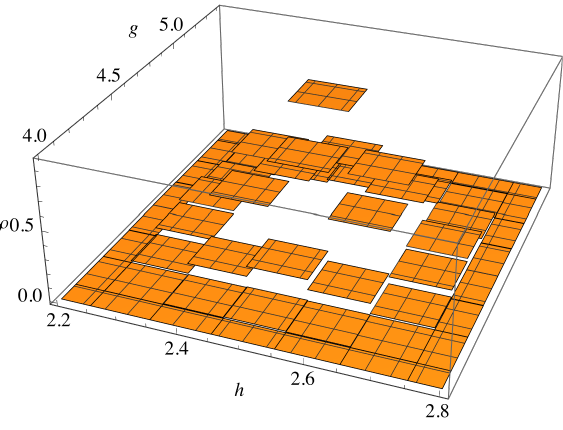}
\caption{The weights $\rho(h,g)$ defined in \Eq{rho(h,g)}, for $h=2.5$ and $g= 4.75$ (off grid).}
\label{f:fit-weights}
\end{figure}

\subsection{Implementation}

Our ED data are such that we can use $N_m=8$ to $17$, i.e.\ a maximal $N_m$ per fit of 
$N_m^{\rm max}=12$ to  $17$. Large sizes are possible since we only need the GS energy. For the order of approximation, we tried $o=2$, $3$, $4$, $5$ and $6$ ($o$ counts the number of terms beyond the constant $E_0$). Larger orders $o$ are not necessarily better, as numerical artifacts are amplified in ways possibly not detectable. Common sense and experience lead us to consider $o=4$ optimal, i.e.\ the  form given in \Eq{half-integer-fit}.  

We now evaluate $\chi$: following  \cite{ZhuHanHuffmanHofmannHe2023}, we plot $h$ on the horizontal axis, and $g$ on the vertical axis. 
We first show a 3D plot, see Fig.~\ref{E3halfoverE-fitorder=4-nmMax=16SqrtFit}, then a heat-plot on Fig.~\ref{E3halfoverE-fitorder=4-nmMax=15SqrtFit-max-curvature-slope}. On the latter, the   critical line of \cite{ZhuHanHuffmanHofmannHe2023} is given in white (partially hidden under blue dots), with a white shamrock marking their {\em sweet spot} (best agreement with Ising CFT). 
The dark blue dots in Fig.~\ref{E3halfoverE-fitorder=4-nmMax=15SqrtFit-max-curvature-slope} mark the valley floor, defined as follows: look at the Hessian $  H_{ij}:=\partial_i \partial_j \chi (h,g)$, where $i,j \in \{h,g\}$. Since $  H_{ij}$ is symmetric, it has two eigenvalues, the curvatures, and two  eigenvectors, which are orthogonal. We take the eigenvector in the direction of the larger curvature, and ask that the slope in this direction vanishes\footnote{This definition   uses the  metric of the coordinates $h$ and $g$. It is not invariant under reparametrization, e.g.\ $\{h,g\}\to \{h,g+h\}$. 
The result is rather similar if we look at a vanishing slope in the $h$ direction only;  this we believe is what \cite{ZhuHanHuffmanHofmannHe2023} did in their optimization procedure. Not knowing the metric will haunt as later, see section \ref{Curvature scaling}.}.

This is a highly non-linear operation on the numerical data generated on a grid with step size $\delta h =1/20$, $\delta g =1/5$, for which we need a smooth interpolation. 
This is obtained by fitting a polynomial of maximal degree four (15 coefficients) to the $6\times 6$ neighbors, weighted by 
\be\label{rho(h,g)}
\rho(h,g):= \exp\left( {-\alpha \bigg[ \frac{(h_i-h)^2}{\delta h^2} + \frac{(g_i-g)^2}{\delta g^2} \bigg]} \right),
\ee
where   
$\alpha=0.6$ is a phenomenological parameter. An example for the weights is given in Fig.~\ref{f:fit-weights}.
Compared to lattice-based approaches which are discontinuous when a new interpolation point enters, our  procedure is very smooth. If the data turn out to be noisy, one can decrease 
$\alpha$ to effectively include more points in the fit. The number of neighbors is chosen s.t.\ additional points   have vanishing weight.

\subsection{Results for the location of the critical point}
As Fig.~\ref{E3halfoverE-fitorder=4-nmMax=15SqrtFit-max-curvature-slope} attests, we can  locate    the phase-transition line of Ref.~\cite{ZhuHanHuffmanHofmannHe2023} which corresponds to the white dashed line. Our valley of $\chi (h,g)$ is marked by blue dots. On this phase-transition line the best agreement with the Ising CFT is achieved at the position of the white shamrock \cite{ZhuHanHuffmanHofmannHe2023}, while the nearby global minimum of $\chi (h,g)$ is marked by a cyan diamond.
\begin{figure}[t]
{\setlength{\unitlength}{1cm}
{\begin{picture}(8.6,5.6)
\put(0,0){\fig{1}{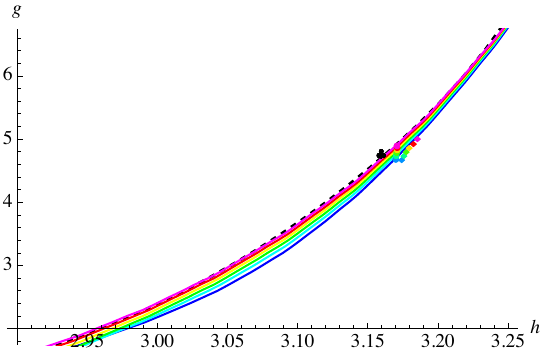}}
\put(0.53,1.6){\fig{0.51}{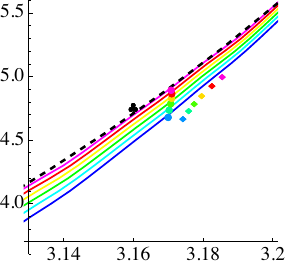}}
\end{picture}}%
}
\caption{Main plot: Dependence on $N_m^{\rm max}$ of the critical curve and sweet spot (dots) from a grid with $\delta g=0.2$, $\delta h=0.05 $; from  $N_m^{\rm max}= 12$ (blue), over $N_m^{\rm max}= 13$ (cyan) to $N_m^{\rm max}= 17$ (magenta); diamonds mark the minimum on a micro-grid with discretization $\delta g=0.05$, $\delta h=0.01 $. The inset shows a blow-up around the sweet spot. Deviations are indicative of errors due to the finite grid size. In black dashed the critical line of \cite{ZhuHanHuffmanHofmannHe2023}, a black trefle marking its sweet spot.} 
\label{f:critical-lines}
\end{figure}
\begin{figure}[t]
\centering
\begin{tabular}{|c|c|c|c|c|c|}
\hline
$N_m^{\rm max}$ & $h$ & $g$ & $\chi_{\rm min}$ & $\chi''_{\rm max}$ &$\chi''_{\rm min}$\\
\hline
12 & 3.17422 & 4.66314 & 0.0056442 &10.97 &0.01907 \\
13 & 3.17587 & 4.72479 & 0.0054353 &12.86& 0.01188 \\
14 & 3.17750 & 4.78087 & 0.0050100 &14.88&0.00780 \\
15 & 3.17965 & 4.84383 & 0.0045505 & 17.01&0.00599\\
16 & 3.18256 & 4.91919 & 0.0041134 &19.27 &0.00496 \\
17 & 3.18553 & 4.99396 & 0.0037134 &21.64  & 0.00441 \\
\hline
\end{tabular}
\caption{Values for the minimum of $\chi$, its location and the two eigenvalues of $\chi''=\partial_{i}\partial_j \chi$ with $i,j = h,g$ for different $N_m^{\rm max}$, obtained on a grid with  $\delta g=0.05$, $\delta h=0.01 $.}
\label{tab1}
\end{figure}
\begin{figure*}
\fig{0.66}{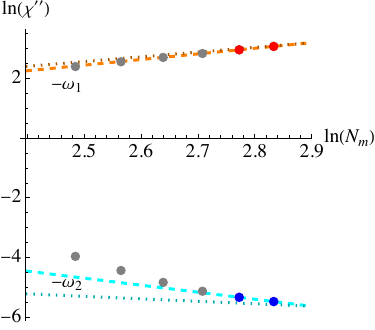}\hfill\fig{0.66}{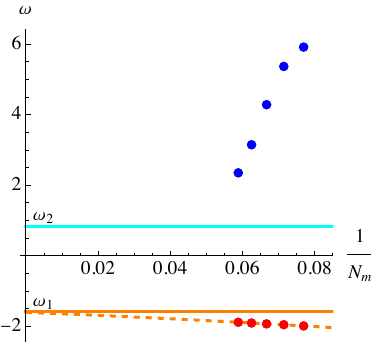}\hfill\fig{0.66}{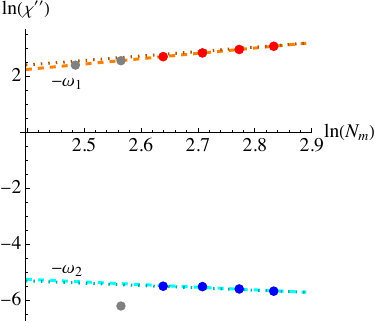}
\caption{Left: Scaling of the two eigenvalues of $\partial_i\partial_j \chi$ (with $i=h,g$) w.r.t.\ $N_m$. The dashed lines are fits with $\omega_1 \approx-1.9$ and $\omega_2 \approx 2.7$, the dotted lines the expected slopes. Gray dots are not included into the fit. Middle: extrapolation of $\omega_1$ and $\omega_2$  to $N_m=\infty$ (using a quadratic polynomial for $\omega_1$). We find $\omega_1 \approx -1.60$ and $\omega_2\approx 1$, to be compared to $\omega_\epsilon = \Delta_{\epsilon}-3 = -1.58738$, and $\omega_{\epsilon'}=\Delta_{\epsilon'}-3=0.82968$ (solid lines) \cite{PolandRychkovVichi2019}. Using a skew transformation to align the last three points for $\omega_2$ (see text) gives $\omega_1$ as before  and $\omega_2 =0.94$, now close to the expected results.}
\label{f:omega-fits}
\end{figure*}
The latter minimum satisfies 
\be\label{CFT-crit}
\hl{\mbox{CFT:~}\qquad \partial_h \chi (h,g) = \partial_g \chi (h,g) = 0.}
\ee
How this minimum depends on the system size  is shown in Fig.~\ref{f:critical-lines}.
There is a small systematic upwards drift  on the minima obtained via interpolation (dots).
We repeated the analysis on a much finer grid (``micro-grid'') with    $\delta g=0.05$ and $\delta h=0.01 $ around the sweet spot, obtaining comparable results, 
see the diamonds on Fig.~\ref{f:critical-lines}, and table \ref{tab1}.

\begin{figure}[t]
\Fig{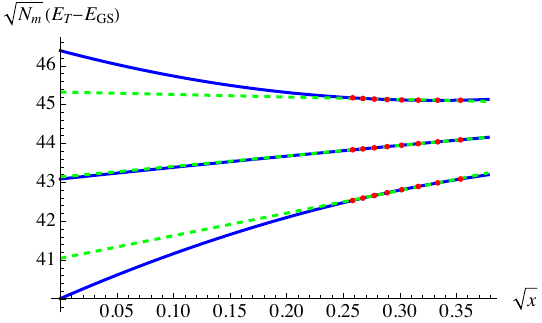}
\caption{The rescaled stress-energy tensor gap $\sqrt{N_m}(E_T-E_{\rm GS})$ ($\ell=2$, $P=Z=1$)  close to the sweet spot: 
$g=4.8$ and $h=3.1$ (bottom), $h=3.2$ (middle), $h=3.3$ (top).
Fits to $\{ 1,\sqrt{x},x\}$ (blue), compared to a  linear fit (dashed green line). Interestingly, the curvature changes sign at the transition.}
\label{f:ETminusEGSoverEGS}
\end{figure}

\begin{figure}[t]
\Fig{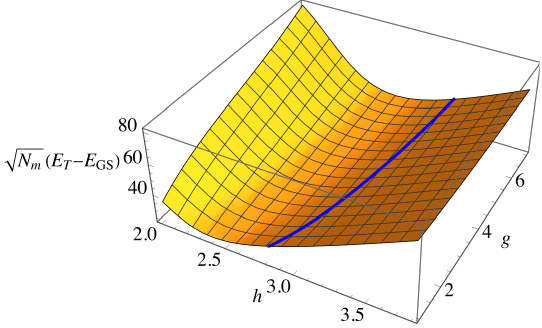}
\caption{The stress-energy tensor gap ($\ell=2$, $P=Z=1$)  multiplied by $\sqrt{N_m}$, and extrapolated to $x=0$. 
 Fit   to $\{ 1,x^{1/2},x\}$, $N_m^{\rm max}=15$.}
 \label{f:ETminusEGSoverEGS2}
\end{figure}%

\begin{figure}[t]
\Fig{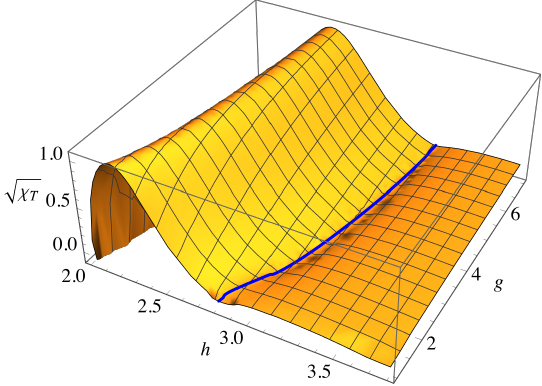}
\caption{$\chi_T $ for  $N_m^{\rm max}=15$; in blue the critical line.  We use the same approach as in Fig.~\ref{E3halfoverE-fitorder=4-nmMax=16SqrtFit} for $\chi$.}
\label{E32overET-E0-Nm=15-case=9}
\end{figure}

\begin{figure}
\Fig{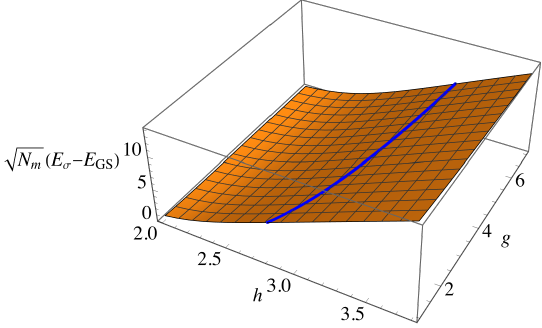}
\caption{$\sqrt{N_m} (E_\sigma - E_{\rm GS})$ for $N_m=15$. Note that the gap vanishes for small $h$: one is in the non-critical ferromagnetic phase.}
\label{Esigma-E0-Nm=15-15-case=9}
\end{figure}

\begin{figure}[t]
\Fig{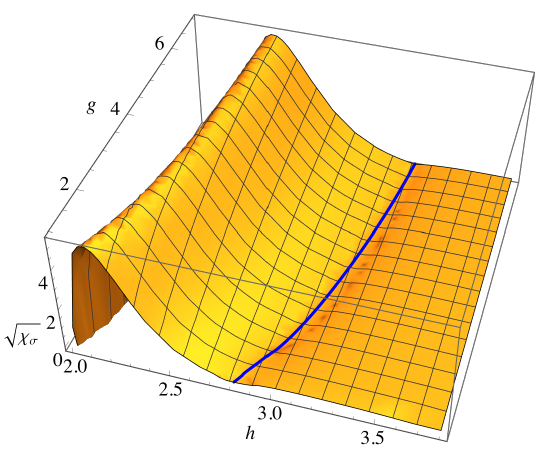}
\caption{$\chi_\sigma $ for  $N_m^{\rm max}=15$; in blue the critical line.  We use the same approach as in Fig.~\ref{E3halfoverE-fitorder=4-nmMax=16SqrtFit} for $\chi$, plotting the signed root $\sqrt{\chi_\sigma} $.}
\label{E32overEsigma-E0-Nm=15}
\end{figure}

\subsection{Curvature scaling} 
\label{Curvature scaling}
\Eq{chi-expansion} states how $\chi$ behaves close to its minimum. Since our numerical data are obtained with two parameters, $\{h,g\}$, 
we should see an approximation involving the two leading parity-even scalars $\epsilon$ and $\epsilon'$:
\bea
\chi   &\simeq& \chi_{\rm min} +   \frac{\lambda_1^2}{N_m^{\omega_1}}+   \frac{\lambda_2^2}{N_m^{\omega_{2}}},    \\
\left( \lambda_1 \atop \lambda_2 \right) &=& \mathbb M  \left( h-h_{\rm c} \atop g-g_{\rm c} \right).
\eea 
The matrix $\mathbb M$ has four parameters. Two can be used for rescaling $\lambda_1$ and $\lambda_2$; since this  only shifts the curves in  plot \ref{f:omega-fits} vertically, we discard it. The next parameter is for  a rotation.
Independent of $N_m$, we find a rotation by $-0.046$ radians  ($-2.7 ^\circ$). 
As a result,  $h-h_{\rm c}$ strongly aligns with $\lambda_1$.  
On Fig.~\ref{f:omega-fits} we show how the largest and smallest eigenvalues  of $\chi'' := \partial_i\partial_j \chi$ with $i,j=h,g$ 
(see Fig.~\ref{tab1}) scale with $N_m$. This allows us to extract $\omega_1 \approx -1.9$, and $\omega_2 \approx 2.7$ from the last two data points. 
We plot the successive slopes  in the middle   of Fig.~\ref{f:omega-fits}. For $\omega_1$ it allows us to extrapolate   $N_m\to \infty$, resulting in 
\be
\omega_1 \approx -1.60, \quad  \omega_\epsilon = \Delta_{\epsilon}-3 = -1.58738 . 
\ee
This is   close to the  numerical bootstrap result \cite{PolandRychkovVichi2019} for $\epsilon$. 

It is more delicate to extract $\omega_2$, as the decreasing slope on the left plot  in Fig.~\ref{f:omega-fits},   shown in the middle plot, asserts. 
We still have one parameter left in  $\mathbb M$: it corresponds to an angle different from $90 ^\circ$ between the microscopic 
coupling constants. 
To exploit this, a simple procedure   is to write $\chi $ to second order as 
\be\label{skew-ansatz}
\tilde \chi =  {\chi''_{\rm max } }\frac {\lambda_1^2}2 + {\chi''_{\rm min } }\frac {\lambda_2^2 + \alpha \lambda_1 \lambda_2}2, 
\ee
 and   redo the scaling analysis of the data in Fig.~\ref{tab1} with  the eigenvalues of $\partial_i\partial_j \tilde \chi$, optimizing $\alpha$ s.t.\ the last three points for $\omega_2$ lie on a straight line, see the right plot on Fig.~\ref{f:omega-fits}. With the optimal $\alpha= 59.868$ we    find   the last four points to align, which allows us to extract 
\be
\omega _2 \approx 0.94, \quad \omega_{\epsilon'}=\Delta_{\epsilon'}-3=0.82968.
\ee
This is now close to the prediction  \cite{PolandRychkovVichi2019} from the numerical bootstrap   for $\epsilon'$. 
As we have exhausted our free parameters, we can not go  further. 

\subsection{Alternative normalizations}
\label{Alternative normalizations}
The alert reader will object that $E_0$ is not universal\footnote{Our procedure continues to work as long as $E_0(h,g)-E_0(h_{\rm c},g_{\rm c})$ has an expansion of the form \eq{14}.}. We propose two ways out of this dilemma, which demand to calculate one more 
eigenvalue, either $E_T$, the subleading contribution in the GS sector corresponding to the stress-energy tensor, or $E_\sigma$, the lowest-lying  parity-odd state.

Let us first consider $E_T$. On Fig.~\ref{f:ETminusEGSoverEGS} we show how weakly $E_T-E_{\rm GS}$ depends on $x$. We found it appropriate  to fit to $\{1, \sqrt x, x\}$. 
Interestingly, the curvature in $\sqrt x$ changes sign at the transition; one should explore this further. 
Fig.~\ref{f:ETminusEGSoverEGS2} shows the resulting extrapolated value of the stress-energy gap as a function of $h$ and $g$. 
This allows us to define 
\be
\chi_T := \frac{E_{3/2}}{\sqrt{N_m}(E_T-E_{\rm GS})} .
\ee
A plot of this function is shown on Fig.~\ref{E32overET-E0-Nm=15-case=9}, which should be compared to Fig.~\ref{E3halfoverE-fitorder=4-nmMax=16SqrtFit}.

A second alternative is to   normalize by the gap of the first parity-odd operator $\sigma$,  
 \be
\chi_\sigma := \frac{E_{3/2}}{\sqrt{N_m}(E_\sigma-E_{\rm GS})} .
\label{chiT}
\ee
While the  $\sigma$-gap is rather insensitive to perturbations  close to the Ising CFT \cite{LaeuchliHerviouWilhelmRychkov2025}, 
it depends strongly on $h$, as Fig.~\ref{Esigma-E0-Nm=15-15-case=9} attests. 
Trying to extrapolate to $N_m=\infty$ even with a linear fit in $\sqrt x$ and two consecutive system sizes leads to a non-sensical negative gap. For this reason, we use only one system size for the normalization in \Eq{chiT}. The result is shown on Fig.~\ref{E32overEsigma-E0-Nm=15}, which should be compared to Figs.~\ref{E3halfoverE-fitorder=4-nmMax=16SqrtFit} and~\ref{E32overET-E0-Nm=15-case=9}. 

\begin{figure}[t]{\setlength{\unitlength}{1cm}
{\begin{picture}(8.6,5.6)
\put(0,0){\fig{1}{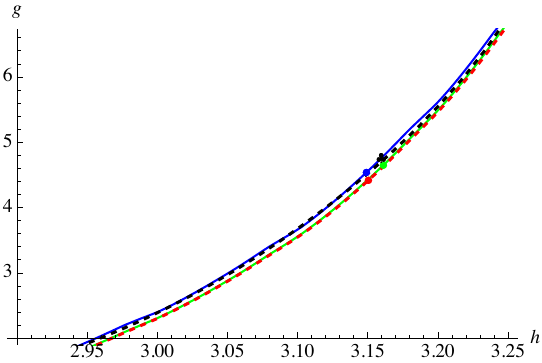}}
\put(0.53,1.6){\fig{0.51}{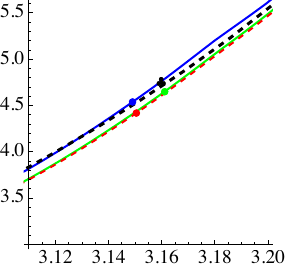}}
\end{picture}}%
}
\caption{As figure \ref{f:critical-lines}, with different denominators at size $N_m^{\rm max}=15$: $\chi$ (green), $\chi_T$ (red dashed) and $\chi_\sigma$ (blue).  In black dashed the critical line of \cite{ZhuHanHuffmanHofmannHe2023}, a black trefle marking its sweet spot.}
\label{Comparison-chi-chiT-chiSigma}
\end{figure}

We   now use $\chi$, $\chi_T$ and $\chi_\sigma$ at $N_m^{\rm max}=15$, and repeat the analysis for the critical line and sweet spot. 
Fig.~\ref{Comparison-chi-chiT-chiSigma} shows this comparison, for    a slightly coarser grid with $\delta g=0.2$, $\delta h=0.1$. The resulting phase-transition lines lie close together, as do their sweet spots; deviations are well below the resolution of the computing grid. 
Our conclusion is that $\chi$, $\chi_T$ and $\chi_\sigma$ give comparable results.

\begin{figure}[t]
\fig{0.49}{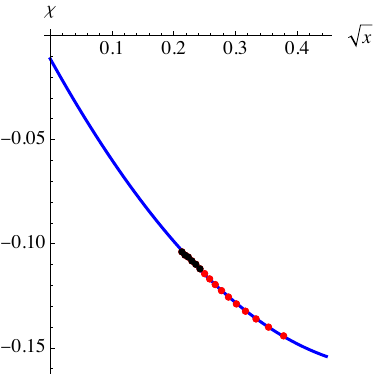}\hfill
\fig{0.49}{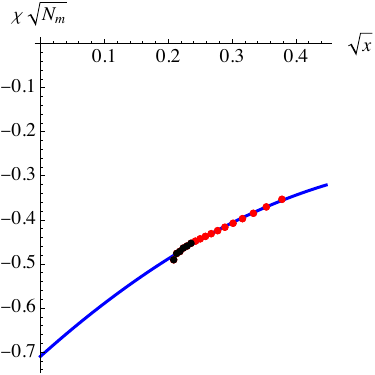}
\caption{Left: $\chi$ as a function of $\sqrt x$ for $N_m$ ranging from $N_m=6$ to $N_m=23$, at the sweet spot of \cite{ZhuHanHuffmanHofmannHe2023}, namely $g=4.75$, $h=3.16$.
The black points are obtained via DMFT, and small numerical errors   increase in the differences we take.
(This is apparent in the last data point).
To the right we show that the extrapolated value is so small, that it is consistent with a different scaling, namely 
that $\chi \sqrt{N_m}$ converges for $N_m\to \infty$.  All fits are quadratic polynomials in $\sqrt x$.
}
\label{discrete-Lap-EGS}
\end{figure}

\subsection{Universality and $\ca F$-function}

The function $ \ca F$ in \Eq{4}   is universal, and recently values for it have been reported \cite{GiombiKlebanov2015,HuZhuHe2024}:
\bea\label{17}
\ca F_{\rm free~theory}&=& \frac{\ln (2)}8 - \frac{3\zeta(3)}{16 \pi^2} \simeq 0.0638071~~ \mbox{\cite{KlebanovPufuSafdi2011}}\nn \\
\ca F_{\rm Ising}^{\rm fuzzy} &=& 0.0612(5) \mbox{~~~(fuzzy sphere)~~ \cite{HuZhuHe2024} }\nn \\
\ca F_{\rm Ising}^{4-\epsilon} &=& 0.0610 \mbox{~~~~~~~~$(4-\epsilon)$~~ \cite{GiombiKlebanov2015}} 
\eea
Can $\ca F$ be accessed  in our approach? 
An obvious  guess is to take $E_{3/2}$ in units of the stress-energy tensor\footnote{The number $d=3$ is   the dimension of the stress-energy tensor.},  
\be\label{19}
  \ca F   \stackrel?=   3 \chi_T .
\ee
At size $N_m=15$, we find at the sweet spot
\be
3 \chi_T \approx  0.00013 \mbox{~~at~~} h = 3.16,~g=4.51. 
\ee
This value is much smaller than $\ca F$ reported in \Eq{17}. 
It seems that in our approach $\chi$ may even vanish, as we tested in Fig.~\ref{discrete-Lap-EGS}:  we evaluated $\chi$ 
at the sweet spot of \cite{ZhuHanHuffmanHofmannHe2023}, using DMRG results for larger systems (the last point at $N_m=23$ may not have converged.) By plotting both $\chi$ and $\chi \sqrt{N_m}$ we see that this is consistent with $\chi=0$ at the 
sweet spot. So either much larger system 
sizes are needed to extract $\ca F$, or $\chi_T$ has no connection to  $\ca F$ and  vanishes at the transition. This may be expected, given that  a quantum system on 
the fuzzy sphere is equivalent to $S_2 \times \mathbb R = \mathbb R^3$, and the latter should not have an anomaly.

\section{Conclusion}
We have shown how the phase-transition line on the fuzzy sphere,  and the sweet spot of optimal conformality can be obtained from a finite-size analysis of the ground-state energy. This is achieved by analyzing the term of relative order $N_m^{-3/2}$ in the ground-state energy, divided by the leading term, as a function of   $N_m$. 
It yields the phase-transition line   of    \cite{ZhuHanHuffmanHofmannHe2023} and their sweet spot with good precision. 
While small system sizes as $N_m=12$  already  allow to locate the phase transition, the precision increases for larger $N_m$. 
A non-negligible advantage of our approach is that it only requires to find the ground-state energy instead of the full spectrum, which is computationally   fast: we need about 10 minutes for $N_m=17$ on  8 cores.

\begin{figure}[t]
\fig{1}{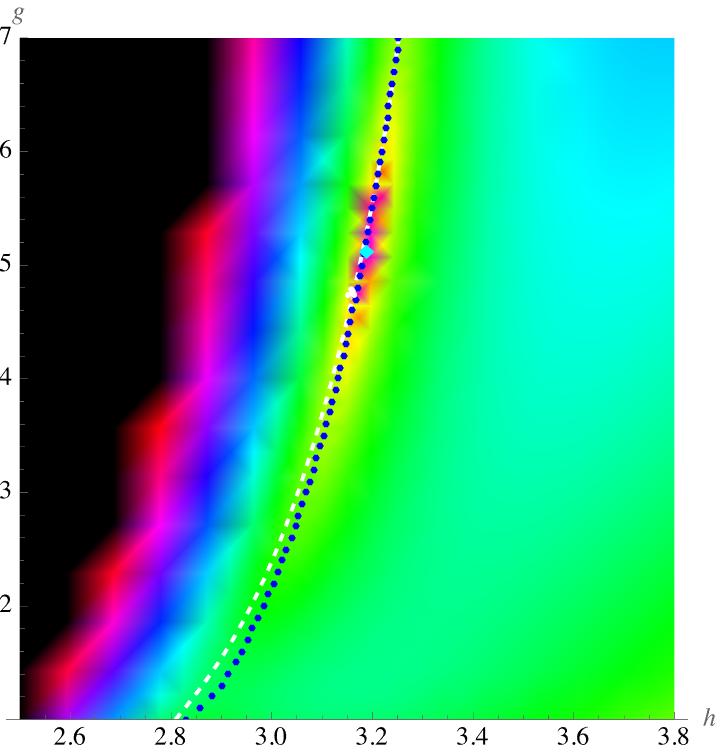}
\caption{The equivalent of Fig.~\ref{E3halfoverE-fitorder=4-nmMax=15SqrtFit-max-curvature-slope}, when 
using a polynomial in $x$ of degree 4 for the fit, and analyzing $-E_{3}/E_0$.}
\label{E3overE-fitorder=4-nmMax=15-case=7-max-curvature-slope}
\end{figure}

An alternative which does not normalize by the ground-state energy is to use either   the stress-energy gap, or the gap of $\sigma$. Both procedures take a universal normalization, give comparable results, but are computationally slightly more costly. 

We hope this procedure may prove useful for locating the critical point in models as the 3-state Potts model in dimension $d=3$, where the transition for real couplings is   first-order 
\cite{Hartmann2005,ChesterSu2022,YangYueTangHanZhuChen2025}.
Suppose we know an approximate location of the minimum in the real plane. We can then approximate $\chi (h,g)$ by a polynomial around this point, and search for solutions 
of \Eq{CFT-crit} in the complex plane, near the approximate minimum.

\acknowledgements
The author thanks Adam Nahum, Junchen Rong and  Slava Rychkov for valuable discussions, and Yin-Chen He, Jesper Jacobsen, Adam Nahum, 
   Zheng Zhou and Wei Zhu for feedback on the draft.

\appendix
\section{Different extrapolation schemes}
\label{s:Alternative fitting schemes}

Since the term $E_{3/2}$ seemingly vanishes at the transition, we can   try to extract the phase-transition line   from 
a fit to a polynomial in $x$. A  reasonable agreement was achieved by considering the term of order $x^4$, as is shown on Fig.~\ref{E3overE-fitorder=4-nmMax=15-case=7-max-curvature-slope}. It finds the phase transition line, as well as the sweet spot, albeit with   less precision. 
While we do not know whether this somehow arbitrary procedure may have a   use, it indicates that the  proposed approach is rather robust.

A final mark of caution:
when using \Eq{half-integer-fit} it is important to not terminate the expansion at $E_{3/2}$, but to keep the two following coefficients $E_2$ and $E_{5/2}$.  Dropping $E_{5/2}$ still allows one to see the phase transition, albeit only at large $N_m$, and with less precision; the sweet spot moves to $h=3.2$, $g=5.3$. Dropping both terms gives a non-sensical result.

\tableofcontents


\ifx\doi\undefined
\providecommand{\doi}[2]{\href{http://dx.doi.org/#1}{#2}}
\else
\renewcommand{\doi}[2]{\href{http://dx.doi.org/#1}{#2}}
\fi
\providecommand{\link}[2]{\href{#1}{#2}}
\providecommand{\arxiv}[1]{\href{http://arxiv.org/abs/#1}{#1}}
\providecommand{\hal}[1]{\href{https://hal.archives-ouvertes.fr/hal-#1}{hal-#1}}
\providecommand{\mrnumber}[1]{\href{https://mathscinet.ams.org/mathscinet/search/publdoc.html?pg1=MR&s1=#1&loc=fromreflist}{MR#1}}

\end{document}